\documentclass[preprint,12pt,3p]{elsarticle}
\makeatletter
\def\ps@pprintTitle{%
 \let\@oddhead\@empty
 \let\@evenhead\@empty
 \def\@oddfoot{}%
 \let\@evenfoot\@oddfoot}
\makeatother

\usepackage{amssymb}
\usepackage{amsmath}
\usepackage[]{hyperref}

\begin{document}

\begin{frontmatter}

\title{Increased Asymmetry of Pit-Over-Peak Statistics\\ with Landscape Smoothing} 

\address[label1]{Department of Civil and Environmental Engineering, Princeton University, USA}
\address[label2]{High Meadows Environmental Institute, Princeton University, USA}

\cortext[cor1]{I am corresponding author}

\author[label1]{Shashank Kumar Anand}
\ead{skanand@princeton.edu}
\author[label1,label2]{Amilcare Porporato \corref{cor1}}
\ead{aporpora@princeton.edu}

\begin{abstract}
The local extremes (i.e., peaks and pits) of the landscape-elevation field play a critical role in energy, water, and nutrient distribution over the region, but the statistical distributions of these points in relation to landscape advancing towards maturity have received limited research attention. In this work, we first explain how the spatial correlation structure of the elevation field affects the counts and frequency distributions of local extremes. We then analyze local extremes statistics for 8 mountainous landscapes worldwide with diverse hydroclimatic forcings and geologic histories using 24 Digital Elevation Models (DEMs) and compare them with complex terrain of the Erythraeum Chaos region on Mars. The results reveal that the spherical covariance structure captures the observed spatial correlation in these cases with the peak frequency distribution agreeing well with the elevation frequency distribution. The ratio of the pit-over-peak (POP) count is linked to the degree to which the pit and peak frequency distributions match, and carries the mark of landscape aging. The relationship between the geomorphic development stage (quantified by the reduced fatness of the slope-distribution tail) and the deviation of POP values from unity in old mountainous landscapes confirms that the evolution towards smoother topographies is atypically accompanied by reduced pit counts typical of organized valley and ridge patterns.
\end{abstract}


\end{frontmatter}



\section{\label{S1} Introduction}

Mountain peaks have intrigued humankind since immemorable times, while the less frequent topographic depressions mostly have captured human attention when filled with water, like lakes and ponds \cite{fowler2011mathematical,price2013mountain}. The small-scale extremes (local maxima or peaks and local minima or pits) of the elevation field affect several ecohydrological and geomorphological processes, acting as hubs for ridge and valley lines and, in turn, influencing the hydrologic partitioning (infiltration, runoff, etc.), local weathering, sediment transport, etc. \cite{moser2007characterization, kiesel2010incorporating,thompson2010role,le2014power}. From an ecological perspective, these points define the ruggedness of the landscape, which is a critical factor in resolving the habitat selection of species \cite{beasom1983technique, reily1999terrain,sappington2007quantifying}.

Mountainous landscapes comprise a quarter of the land surface. After formation, their relaxation phase is regulated by the interaction of erosion and sedimentation \cite{hack1960interpretation, tucker1998hillslope, willett2002steady, bonetti2017dynamic}. The combined effect of these transport mechanisms, due to the inherent direction induced by gravity, acts differently on peaks and pits (as well as on ridges and valleys), so that old mountain ranges in the later stage of the geomorphic development (e.g., the Smokies) not only have smoother hillslope profiles but also have different pit-over-peak statistics, with some pits either filled or worn away, as opposed to young ranges like the Alps. The effect of landscape aging on extreme statistics is expected to be affected by land-surface properties (vegetation cover, soil properties, etc.), underlying geology, climatic forcing, and anthropogenic disturbances as well \cite{hooke2000history,brecheisen2019micro, bonetti2019effect}. The precise description of these observations, as well as a quantitative link between perceived trends of these topographic features and the geomorphological development stage, is largely unexplored in the literature and forms the motivation of this study.

Here we focus on quantifying and analyzing the difference between the statistical distributions of local extremes (peaks and pits) in diverse mountainous landscapes. We first show using idealized synthetic landscapes that the spatial correlation structure of the elevation field holds the first-order information about the counts and distribution shapes of local extremes. While in these idealized cases the impact of the hillslope erosion and sedimentation with the inherent directionality due to gravity is absent, a significant difference in pit and peak distributions may exist in natural landscapes.

To quantitatively evaluate these differences, we define the pit-over-peak ratio (POP) for 24 Digital Elevation Models (DEMs) from 8 mountainous landscapes over the earth (and one from the Erythraeum Chaos region on Mars). The deviation of POP from unity shows the decrease of local pit counts compared to the peaks and explains the dissimilarity between pit and peak distributions. Comparison of POP values with the fatness of the slope distribution tails hints at the possible linkage between the geomorphic development stage and the local extremes statistics. This result indicates that peak and pit distributions evolve asymmetrically as the hillslopes get smoothened by the interplay of erosion and sedimentation during the relaxation phase of the mountainous landscapes.

\section{\label{S2} Spatial Correlation and Extremes in Synthetic Landscapes}

We begin by studying how the spatial correlation structure influences the counts and frequency distributions of local extremes (i.e., peaks and pits) for synthetic landscapes by employing both 1D and 2D numerically generated elevation fields. 

\subsection{\label{SS21} 1D Elevation Profiles}

For 1D elevation profiles, the homogeneous elevation field ($z$) was characterized by standard Gaussian probability distribution with $\sigma$ as the standard deviation. The frequency distribution of peaks ($p^+\left(z\right)$) can be obtained analytically for this case\cite{soong1973, vanmarcke2010random} as
\begin{equation}
    p^+\left(z\right) = \frac{\sqrt{1-{\nu}^2}}{\sqrt{2 \pi \sigma^2}} \exp \left(\frac{-z^2}{2 {\sigma}^2 \left( 1-{\nu}^2 \right) }\right) + \frac{\nu z}{2 {\sigma}^2} \exp \left( \frac{-z^2}{2 {\sigma}^2}\right) \left[ 1 + {\rm erf} \left( \frac{\nu z}{\sigma \sqrt{2\left(1-{\nu}^2 \right)} }\right)\right],
    \label{max_the_dis}
\end{equation}
where ${\rm erf} \left(\cdot \right)$ is the error function \cite{zill2011advanced} (a detailed derivation is provided in the Supplementary Material). The parameter $\nu$ represents the ratio of average number of up-crossing (positive slope) above the (zero) mean elevation to the expected number of local maxima/peaks per unit length. Using the definition of spectral moments ($\lambda_k = \int_{-\infty}^{\infty}|\omega|^k \mathbf{S_\omega} d\omega$, with $\omega$ as the associated frequency and $\mathbf{S_\omega}$ as the spectral density function), $\nu$ can be written as 
\begin{equation}
    \nu = \sqrt{\frac{{\lambda_2}^2}{\lambda_0 \lambda_4}}.
\label{nu_eq}
\end{equation}
The limiting case, $\nu \rightarrow 1$, can be interpreted as there is one likely peak on average for each zero up-crossing, resulting in a Rayleigh distribution. For the other limit $\nu \rightarrow 0$, we can expect a very high number of peaks for every up-crossing, giving the same Gaussian distribution of the elevation field. As equation (\ref{nu_eq}) indicates, the value of $\nu$, as well as the count and frequency distribution shapes of local extremes, is controlled by the dispersion or spread of the spectral density function. This information in the frequency domain can also be translated into the spatial domain as spatial correlation and spectral density are related to one another (Fourier Transform pairs), suggesting that different correlation structures of the elevation field modify the frequency distributions of local extremes in the domain \cite{stoica1997introduction}. 

\begin{figure}[!hbt]
\centering
\includegraphics[width=\linewidth]{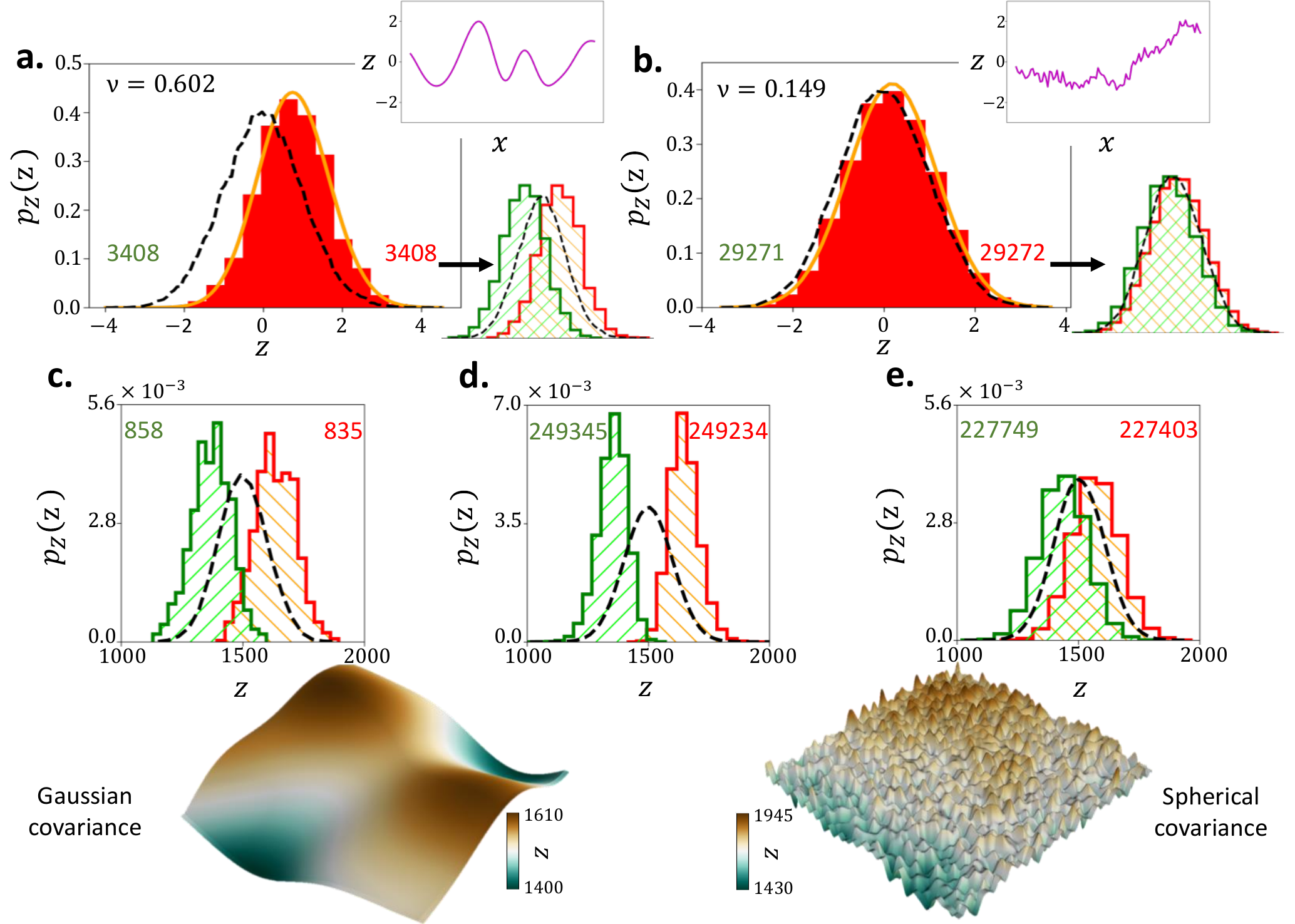}
\caption{ \label{fig:one} Effect of the covariance structure on the extremes distributions for 1D/2D homogeneous elevation field. (a,b): 1D elevation field of $10^5$ points with unit spacing derived from standard Gaussian distribution for Gaussian/spherical covariance structure with 10/100 units length-scale. The frequency distribution of the elevation is displayed as a black-dashed curve. Numerically obtained peaks distribution is shown as a red-filled step graph compared to the (orange) theoretical distribution. The inset in each panel shows the 100 units of the 1D field regarding the corresponding realization. The distribution of (green-hashed) pits is also compared with (red-hashed) peaks frequency distribution. (c,d,e): 2D elevation field realized from the Gaussian distribution (square side = 1500 units, mean = 1500 units, and standard deviation  = 100 units). (d): Spatially uncorrelated elevation field. (c,e): Using isotropic Gaussian/spherical covariance structure with 20/100 length scale. The elevation field frequency distribution is displayed as a black-dashed curve with peaks as red-hashed step graph, and pits as green-hashed step graph with counts mentioned for each case. The inset in each panel shows a 3D surface plot (from a square portion with a 50 unit long side) of the corresponding realization.}
\end{figure}

In this analysis, we focused on the Gaussian and spherical covariance structures, which have contrasting forms near small lag/spatial separation. The Gaussian covariance structure as a function of lag distance has a (twice-differentiable) parabolic shape at the zero-lag (origin), ensuring short-scale spatial continuity for nearby elevation values. This short-scale spatial continuity is absent from the spherical covariance structure that starts linearly from the zero-lag distance, forming a cusp at the origin \cite{pyrcz2014geostatistical}. A 1D elevation field for both cases was realized, where peaks/pits were categorized as the points in the series having values higher/lower than two adjacent neighbors. Results for the Gaussian covariance structure with the length scale of 10 units and the spherical covariance structure with the length scale of 100 units are shown in figure \ref{fig:one} (a,b). The black curve displays the frequency distribution of the elevation field and the red-filled step graph shows the numerically obtained distribution of the peak matching with the orange curve for the analytical expression given by equation (\ref{max_the_dis}). 

For the Gaussian covariance structure, the peak frequency distribution does not match the elevation frequency distribution, as peaks are largely found at high elevation values (figure \ref{fig:one}(a)). For the spherical covariance structure, a good match in the shape of the peak distribution (analytically and numerically) and the frequency distribution of the elevation field occurs. The pit frequency distributions for both covariance structures appears as a reflection of the respective peak distribution across the $y-$axis  (insets in figure \ref{fig:one}(a,b)). For the Gaussian covariance structure, peak and pit distributions remain right and left of the elevation frequency distribution, while three distributions match for the case of the finite length scale of the spherical covariance structure. Another crucial piece of information is about the peak and pit counts (shown in red and green for both cases), which are high for spherical structure as opposed to the Gaussian covariance structure. This confirms that changing the correlation structure from the Gaussian to spherical not only alters the distribution of pits and peaks but also increases the number of local extremes found in the domain. 

The contrasting effect of two covariance structures on the extreme counts and the shape of the frequency distributions of local extremes can be explained by the computed value of $\nu$ in both cases. For the Gaussian covariance structure, the number of peaks and the number of up-crossing above the mean elevation are reduced by the same proportion so that the value of $\nu$ stays high ($=0.602$) and the peak distribution does not shift leftward. In the case of the spherical covariance structure, the number of peaks remains high but the number of zero-crossing gets reduced such that the value of $\nu$ declines ($=0.149$), and the frequency distribution of peak shift and match with the elevation frequency distribution. These alterations in up-crossing and extreme statistics can be discerned from the spatial fluctuation patterns shown in the inset for both covariance structures.

\subsection{\label{SS22} 2D Elevation Profiles}

We further examined the frequency distributions of peaks and pits for the 2D isotropic field on a raster grid for the same two sets of covariance structures, where the nodes in the domain having values higher/lower than the eight adjacent neighbors were determined as local peaks/pits. 2D elevation field was realized from the Gaussian distribution with 1500 units mean and 100 units standard deviation on a square domain with 1500 units side length. We simulated a base case of elevation field with no spatial correlation to be compared with the Gaussian covariance structure for a length scale of 20 units and the spherical covariance structure for a length scale of 100 units (shown in figure \ref{fig:one}(c,d,e)). The black dashed curve shows the frequency distribution of the elevation field and the red-hashed/green-hashed histogram displays the obtained frequency distribution of peaks/pits.

In the uncorrelated field, the peak/pit distribution is found on the right/left side of the elevation frequency distribution (figure \ref{fig:one}(d)). The count and distributions of peaks and pits for the Gaussian and spherical covariance structures reflect the findings of the 1D case above. Shapes of the peak and pit frequency distributions do not shift for the Gaussian covariance structure with a reduction in the counts of both extremes (figure \ref{fig:one}(c)). The spherical covariance structure shifts both peak and pit distributions towards the frequency distribution of the elevation field with the high number of peak and pit counts in this scenario (figure \ref{fig:one}(e)).  The shifts in the spatial fluctuation patterns due to two sets of covariance structures can be visualized from the 3D surface plots of the square domain (50 unit long side) taken from the realized elevation fields. As the insets in panels (c,e) of figure \ref{fig:one} show, a small number of peaks/pits occur mostly above/below the mean elevation value for the Gaussian covariance structure, unlike a high count of peaks/pits scattered both above and below the mean elevation field with three frequency distributions matching for the case of spherical covariance structure. 

\section{\label{S3} Real Landscapes}

\begin{figure}[ht]
\centering
\includegraphics[width=\linewidth]{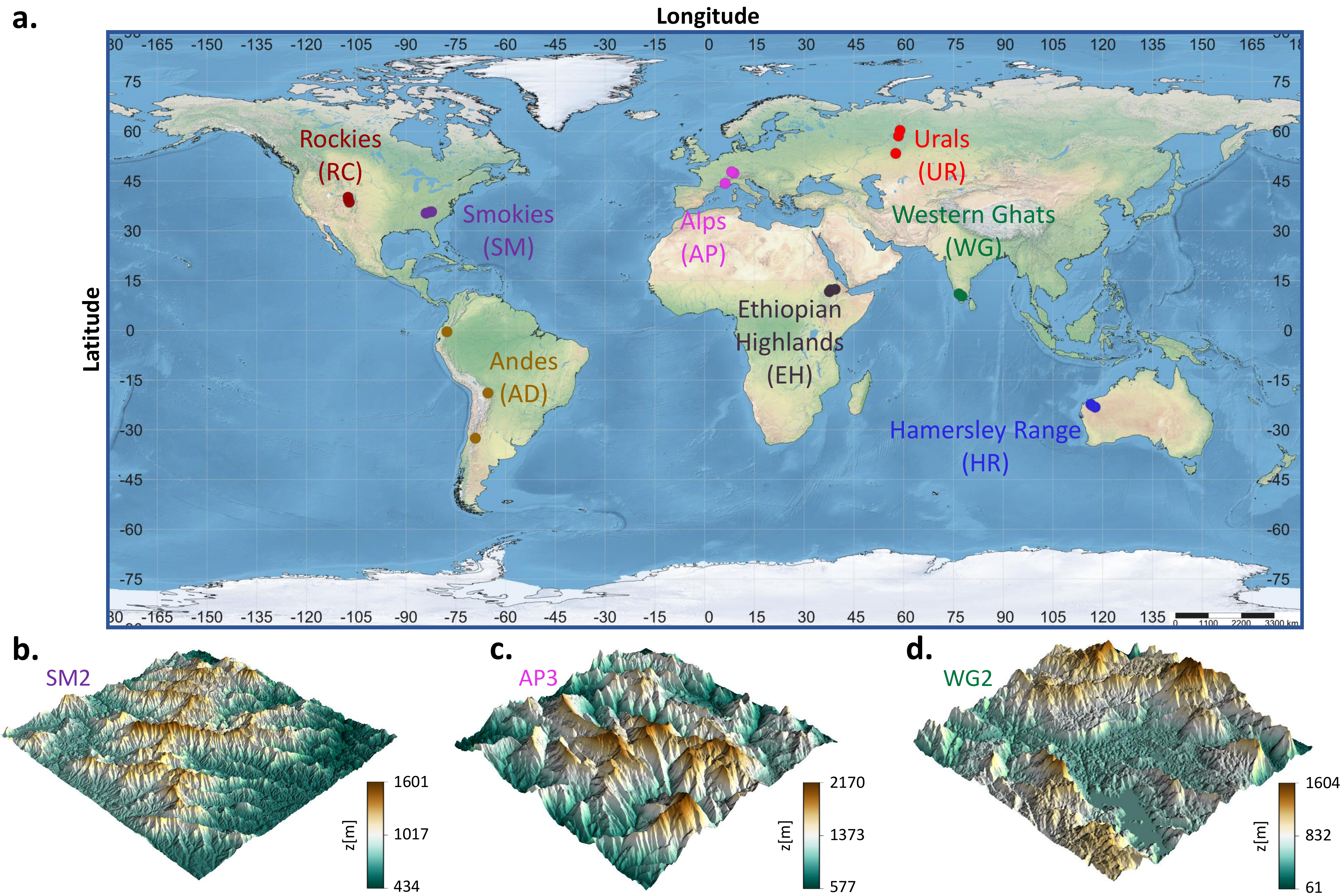}
\caption{ \label{fig:two} (a): Locations of 24 DEMs around the globe used in the study (3 DEMs per mountainous range). Three examples of 3D surface plots of the elevation field for DEMs (b): Smokies (SM2), (c): Alps (AP3), and (d): Western Ghats (WG2).}
\end{figure}

Informed by the theoretical and numerical results of the previous section on the role of the spatial correlation structure on the local extremes, we examined the extreme distributions of the elevation field for different mountain ranges (Figure \ref{fig:two}). Besides the correlation structure impacts, we also wanted to explore the potential effect of hillslope smoothing on these local extremes statistics. We employed three distinct classes of Digital Elevation Models (DEMs) for this analysis - 3D Elevation Program DEMs and ASTER Version 3 DEMs for 24 DEMs across the globe, DEM by NASA Pacific Regional Planetary Data Center for one DEM in the Erythraeum Chaos region on Mars \cite{us20171, asterv3, malin2007context}. This was done to lessen the bias of the particular type of sensor errors and the applied data-processing algorithm on the final constructed DEMs. The reader is referred to the Supplementary Material for more information on the types of data sets used and the pre-processing performed before the statistical analysis.

\subsection{\label{SS31} Pit vs Peak Frequency Distributions}
\begin{figure}[!hbt]
\centering
\includegraphics[width=\linewidth]{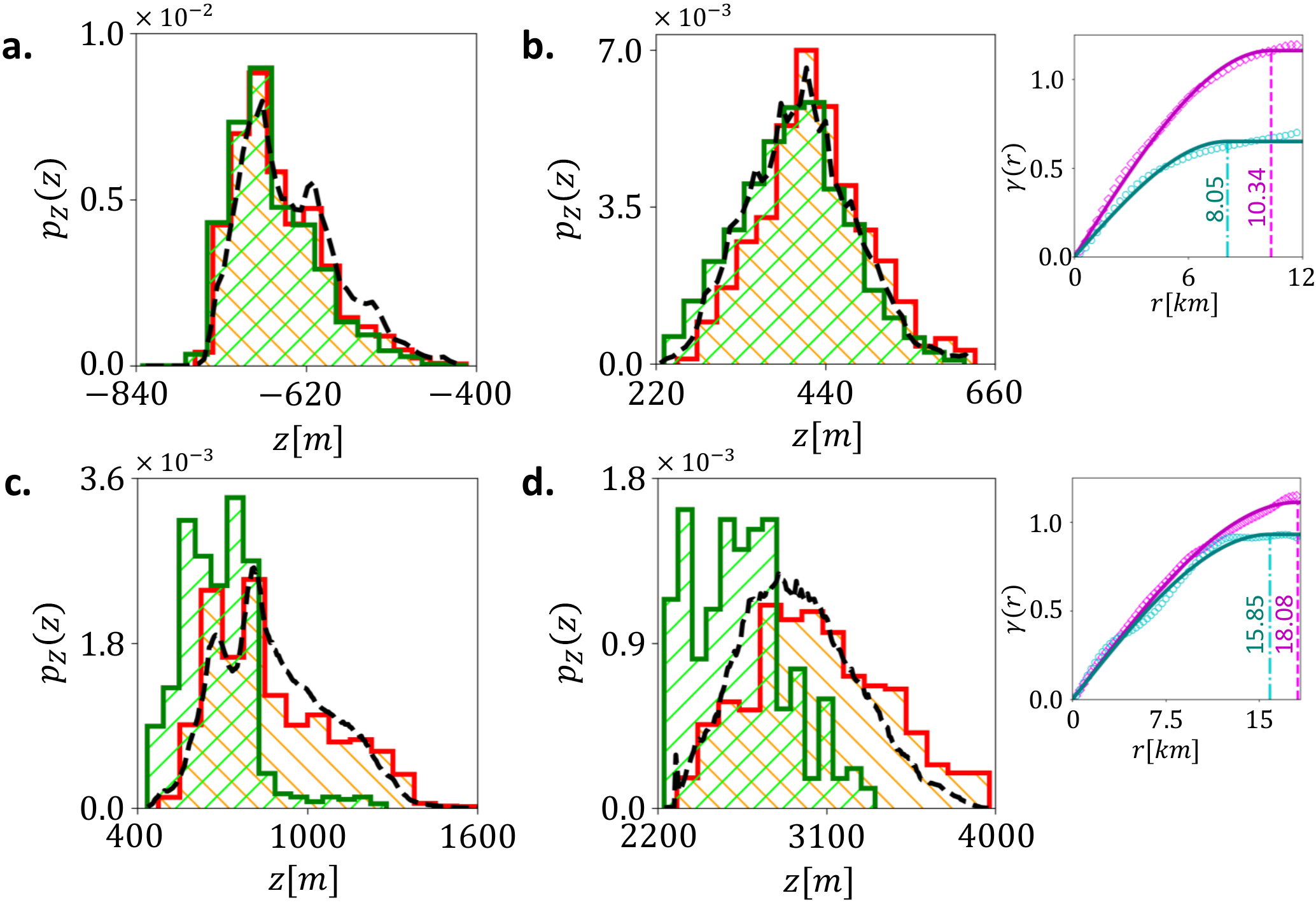}
\caption{ \label{fig:three} Plots of the frequency distributions of elevation field (black-dashed curve), peak (red-hashed step graph), and pit (green-hashed step graph) for (a): Mars (MA), (b): Urals (UR2), (c): Smokies (SM2), and (d): Rockies (RC3). Panels (a,b) correspond to the cases where a good match among three frequency distributions is observed. Panels (c,d) display the cases where elevation field distribution matches the peak frequency distribution, but not the pit frequency distribution. The variogram as a function of lag distance $r$ along $x-$ (magenta) and $y-$ (teal) axes for cases (b) and (d) are shown with vertical dotted lines indicating the length scale (the number of grid cells multiplied by the nominal resolution of 30 m) for the fitted spherical covariance structure along each axis.}
\end{figure}

For all the DEMs considered in this study, we found a good agreement between the frequency distributions of the peak and the elevation field, showing an effectively decoupled functional relationship between peak frequency distribution and hydro-climatic conditions. For pits, we witnessed several cases where they too obey the elevation frequency distribution, with all three frequency distributions matching with each other, but there were mountainous landscapes where the pit frequency distributions do not reflect the same shape as peak and elevation frequency distributions. Between these two observed limits of the frequency distribution ensembles, a few landscapes were observed to maintain some similarities between pit frequency distributions with elevation field and peak frequency distributions. We juxtapose two limits of frequency distribution ensembles (peak, pit, and elevation) in figure \ref{fig:three}, where the frequency distribution of the elevation field is shown as a black-dashed curve, peak frequency distribution is the red-hashed graph, and pit frequency distribution is displayed as a green-hashed graph. Panels (a) and (b), from Mars and Urals, present the landscapes where three distributions follow the same shape. Panels (c) and (d) display the DEMs from Smokies and Rockies, where the elevation field distribution agrees with peak frequency distribution but differs from the pit frequency distribution.

We examined the spatial correlation of the elevation field for selected DEMs by estimating the variogram and performing the covariance structure fit \cite{sebastianmuller2021}. The results for two DEMs are presented in the insets of figure \ref{fig:three}(b,d); UR2 (three frequency distributions match) and RC3 (peak and elevation frequency distributions coincide, different from the pit distributions). The variograms of the elevation field as a function of lag distance along the $x-$ and $y-$axes are displayed by magenta- and teal-colored points, respectively. The spherical covariance structure fits well for both DEMs along two axes as shown in solid curves and captures well the spatial structure of the elevation field with correlation coefficient $r = 0.997$ for the $x-$axis and $r = 0.988$ for the $y-$axis in RC3, and $r = 0.997$ for the x-axis and $r = 0.983$ for the $y-$axis in UR2 DEM.

\subsection{\label{SS32} Pit-Over-Peak Ratio (POP)}

The spherical correlation structure of mountainous landscapes elucidates the overlap between the peak and elevation frequency distribution (similar to the synthetic landscapes). However, it does not explain the dissimilarity between elevation and pit distributions observed in some cases. To explore the reasons for this, we examined the pit-over-peak ratio (POP), i.e., the ratio of pit counts to peak counts. POP is around one for synthetic surfaces, with both local extremes frequency distributions resembling the elevation frequency distribution (Figure \ref{fig:one} (e)).

\begin{figure}[!hbt]
\centering
\includegraphics[width=\linewidth]{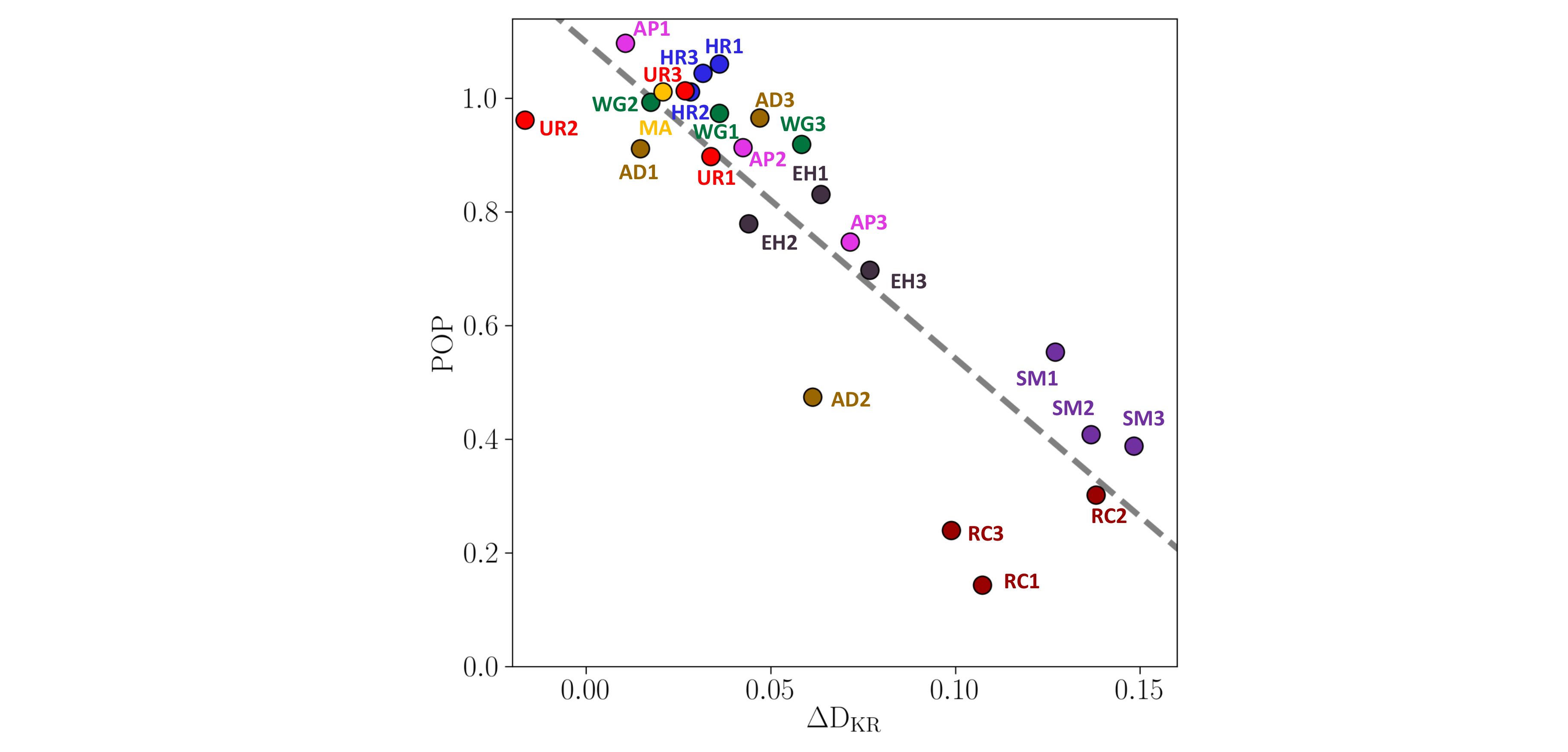}
\caption{\label{fig:four} Categorization of mountainous landscapes using the POP value for the observed matching level of pit frequency distribution compared to the elevation and peak frequency distributions. $x-$ axis presents $\Delta {\rm D}_{{\rm KR}}$ and $y-$ axis shows the POP ratio. The top-left region includes DEMs where the number of peaks and pits (as well as their frequency distributions) match. The bottom-right region includes DEMs where a mismatch in the counts and frequency distributions of peaks and pits are observed. The linear relationship between is shown as gray line (correlation coefficient $r= 0.858$).}
\end{figure}

We quantified the differences between shapes of frequency distributions using Kantorovich–Rubinstein (KR) distance function, which represents the cost associated with the most optimal mapping that shapes one probability distribution into another \cite{panaretos2020invitation}. For 1D distributions, this metric equals the area between cumulative distributions of two frequency distributions \cite{villani2003topics}. We defined $\Delta{\rm D}_{\rm KR}$ as a measure of dissimilarity of pit frequency distribution compared to peak and elevation frequency distributions. The metric $\Delta{\rm D}_{\rm KR}$ was computed by subtracting the KR distance of peak and elevation frequency distributions from the KR distance of pit and elevation frequency distributions, with a lower value where three distributions match as opposed to the higher value of $\Delta{\rm D}_{\rm KR}$ for cases where pits distribution does not conform with the other two frequency distributions (refer to the Supplementary Material for more information).

The relationship between POP and $\Delta{\rm D}_{\rm KR}$ is shown in Figure \ref{fig:four}, where each circle symbol represents a DEM considered in this study. A strong negative correlation between these two variables indicates that as POP reduces, the distance of pit-frequency distribution increases from the elevation and peak frequency distributions. The DEMs in Figure \ref{fig:three} can be mapped to the different regions of the phase space based on POP and $\Delta{\rm D}_{\rm KR}$. Panels (a,b) of Figure \ref{fig:three} (MA, UR2) correspond to the top-left region of Figure \ref{fig:four}, where the count ratio of the pit to the peak is close to unity, and consequently, a good match among peak, pit, and elevation frequency distributions is observed. For DEMs in the bottom-right region (for example, SM2 and RC3 from Figure \ref{fig:three}), there is a mismatch between the pit frequency distribution and the peak and elevation frequency distributions. DEMs like AP3, EH3 belong to the middle region, where the shape of pit frequency distribution only partly agrees with peak and elevation frequency distributions. This analysis confirms that the removal of pits, as opposed to the relatively robust presence of peaks, is the cause of the difference between pit and peak statistics in different mountainous landscapes.

\subsection{\label{SS33} Loss of Pits and Smoothing of Landscapes with Age}

The behavior observed in the previous section has revealed a landscape evolution characterized by a decline in pits while leaving peak statistics relatively unchanged. This asymmetric evolution of extremes likely arises due to the directionality of sediment transport mechanisms set by gravity. In fact, the interplay of erosion and sedimentation processes going from the top of the ridges to the base of the hillslopes and the related valley aggradation influence local pits and peaks differently. Local pits define flow-convergence zones of the elevation field as opposed to local peaks where the flow diverges. Consequently, considering the so-called stream-power law for fluvial erosion, $K_e a^m |\nabla z|^n$ ($a$ is the specific contributing area, $K_e$ erosion coefficient, and $m,n$ positive constants \cite{whipple1999dynamics,royden2013solutions,bonetti2020channelization}), it appears that erosion is more effective in eroding away a pit surrounding given the larger contributing area as opposed to peak erosion where the contributing area is zero \cite{cayley1859xl,florinsky2016digital,bonetti2018theory}. Moreover, at the base of the hillslopes, both colluvial (mass movement of sediment due to gravity) and alluvial (overland or channelized water flow under the action of gravity) processes supply and re-deposit the material from the upper parts of the hillslopes with high chances of filling a local topographic depression (pit).

Several investigations following the study by Strahler \cite{strahler1956quantitative} described the relationship between the statistical properties of the landscape (elevation, slope, gradient, etc.) and its systematic progress towards maturity \cite{vico2009probabilistic, hurst2013hillslopes}. Particularly, the slope frequency distribution parameters were shown to provide a compact metric of the aging while comparing landscapes in different stages of the relaxation phase \cite{bonetti2017dynamic, bonetti2019effect}. These works explained that the landscape age tends to be linked to the slope ($S$) distribution tails via a power law as
\begin{equation}
    p(S) \propto S^\beta,
\label{tailpower}
\end{equation}
for large values of slope in the distributional tails. The exponent $\beta$ is typically higher for young ranges but decreases with age and reaches lower values for old landscapes with smoother hillslopes in the relaxation part of the geomorphic evolution \cite{bonetti2017dynamic,bonetti2019effect}. 

Building on these results, we calculated the exponent $\beta$ of the slope distribution tails for DEMs considered in this study to investigate the modified pit statistics in relation to landscape smoothing with age. Linear regression of the slope distribution tail on the logarithmic scale was performed to compute the exponent value for each DEM. The power-law region for the tails of the slope distributions were considered to extend between distribution values $p(S) = 10^{-0.5}$ and $p(S) = 10^{-2.5}$ (see Supplementary Material for more information). The value of the exponent is rather insensitive to the specified limit values as long as the linear regression is employed to the region of the slope distribution tail that obeys the power-law form \cite{bonetti2017dynamic}.

\begin{figure}[!hbt]
\centering
\includegraphics[width=\linewidth]{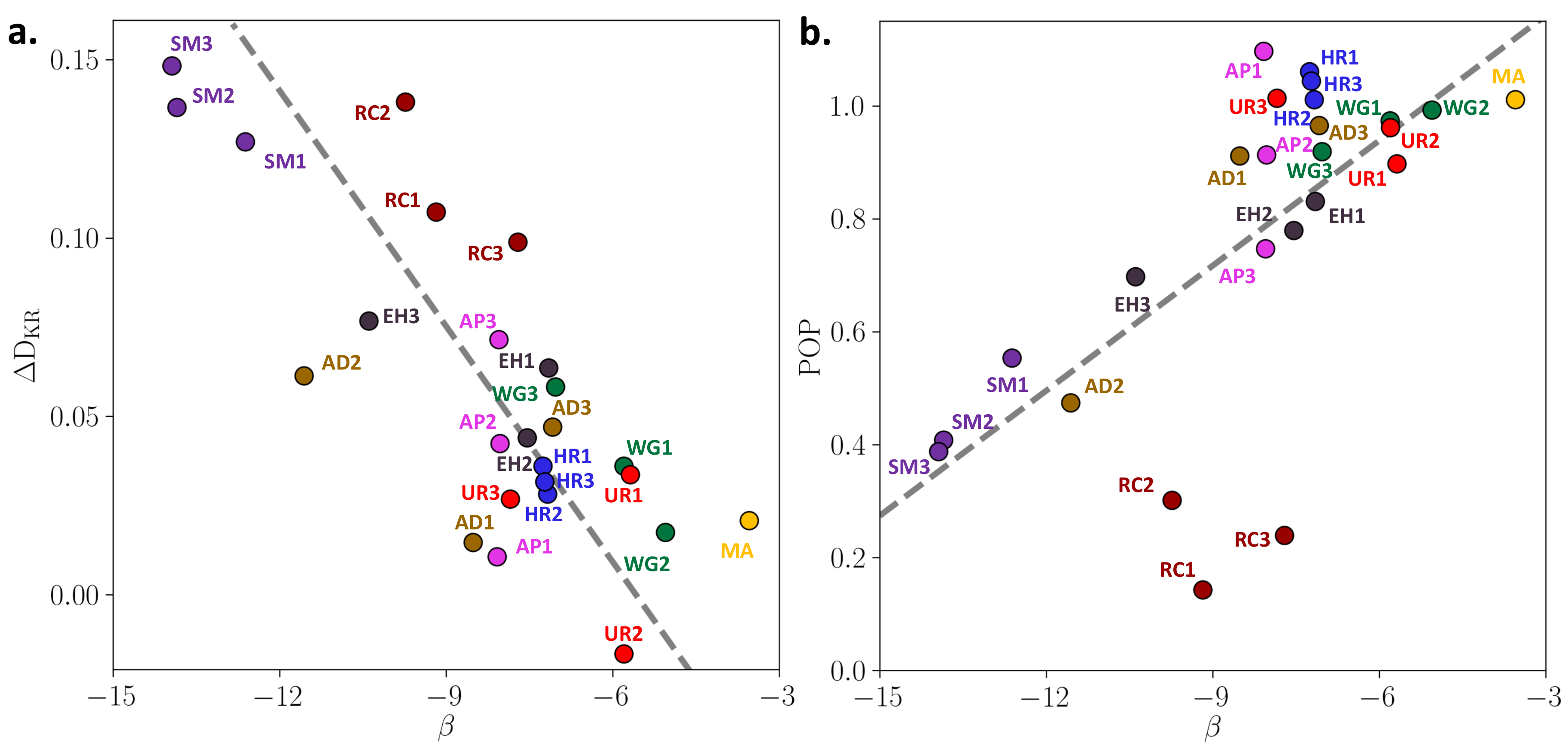}
\caption{\label{fig:five} The linkage between local extremes statistics and smoothing of hillslopes. (a): Plot of $\beta$ and $\Delta{\rm D}_{\rm KR}$ for selected DEMs with the linear trend shown as a gray line (correlation coefficient $r= 0.781$). (b): Space formed by $\beta$ ($x-$ axis) and POP ($y-$ axis) with the linear trend shown as a gray line (correlation coefficient $r= 0.662$). The observed relationship indicates that younger landscapes (large $\beta$ with fatter tails of slope distribution) tend to have a higher degree of similarity of pit distribution shape with peak and elevation frequency distributions with POP value closer to one. The statistical distance of the pit distribution from peak and elevation distributions increases with a reduction in their counts unlike local peaks as landscapes age and reach smoother geometries.}
\end{figure}

For the studied DEMs, we compared metric $\Delta{\rm D}_{\rm KR}$ with the corresponding value of $\beta$. The results are shown in Figure \ref{fig:five}(a), where each symbol represents the position of a DEM in the space formed by $\beta$ ($x$-axis) and $\Delta{\rm D}_{\rm KR}$ values ($y$-axis). This analysis presents a strong correlation between the reduced slope-tail fatness and increased distance of the pit distribution from peak and elevation frequency distributions (gray line showing the linear trend). A similar comparison of POP values with $\beta$ for 25 DEMs is presented in Figure 5(b), where the data displays the connection between the reduction in pit counts relative to the local peak counts as landscapes attain smoother topography. Although there exists a scatter in the trend, it still reveals an evident agreement between modified pits statistics and the smoother hillslope profiles as landscapes age. The presented results extend the linkage between the landscape development stage and the behavior of slope distribution tails to the enhanced asymmetry of local pit-over-peak statistics. As landscapes evolve over time, there is a systematic progression of extreme statistics beyond the reduced hillslope steepness with pits getting reduced, unlike robust peaks, which increases the statistical distance between the two distributions.

Landscapes positioned in different parts of this space have distinctive hillslope morphology as well as pit-over-peak statistics. For example, DEMs like SM1, SM2, and SM3 are situated on the west side of the Great Smoky Mountains with a humid climate. In these ranges, hillslopes have been eroded over the past $\sim 360$ million years \cite{hibbard2000docking}, resulting in gentle rolling ridges and valleys (average $\beta = -13.47$) with fewer pit counts in the region (average ${\rm POP} = 0.45$) and a high degree of mismatch between pit and peak frequency distributions (average $\Delta{\rm D}_{\rm KR} = 0.143$). On the contrary, DEMs like HR1, HR2, and HR3 with ${\rm POP} \approx 1$ belong to the north-west rugged part of Hamersley Range from Western Australia have high topographic unevenness with thin ridges and steep hillslopes (average values of $\beta =-7.22$, $\Delta{\rm D}_{\rm KR} = 0.036$). Local studies confirm these features while describing the region as mountainous deserts with an arid climate \cite{byrne2017,van2017vegetation,gwreportwa2016}. Consequently, less favorable conditions for sediment transport and smooth hillslope morphology have resulted in unaltered pit-over-peak statistics.

The correlation of reduction of pit counts with age also suggests that this process is associated with the so-called `relaxation phase' of landscape evolution, which follows the faster `freezing phase' \cite{sinclair1996mechanism, banavar2001scaling}. During the freezing time phase, the rapid development of the drainage network happens after which its form is reasonably fixed. Once the initial rapid network-forming phase of an evolving landscape passes, the subsequent slow relaxation phase is dominated by the gradual smoothing of hillslopes \cite{fernandes1997hillslope, bonetti2017dynamic}. The interaction of erosion and sedimentation controls the hillslope evolution towards quasi-equilibrium forms in this phase as long as underlying geological and hydroclimatic conditions remain the same \cite{hack1960interpretation, kirkby1971hillslope}. As landscape systematically progresses through this stage towards maturity, asymmetric evolution of pit-over-peak statistics is likely to emerge by the sediment transport mechanisms with reduced hillslope steepness and smoother morphology.

\section{\label{S4} Conclusions}

Statistical analysis of extremes of the elevation field performed in this study shows the enhanced asymmetry of local pit-over-peak statistics as mountainous hillslopes smooth and advance towards geomorphic maturity. We first determined that the spherical covariance structure captures well the spatial correlation of the elevation field in complex mountainous landscapes. The covariance structure explains a good match between peak and elevation frequency distribution shapes, a feature observed in most mountainous landscapes. 

The matching degree between pit frequency distribution shapes compared to peak and elevation frequency distributions varied for landscapes analyzed in this study. To explain these changes in pit distributions quantitatively, we defined a metric POP providing the ratio of pit counts to peak counts. Its relationship with $\Delta{\rm D}_{\rm KR}$ (a measure of dissimilarity of pit distribution to peak and elevation frequency distributions) reveals that the reduction in local pit counts caused the mismatch between frequency distribution shapes. 

During the long time-scale of relaxation phase, the peak statistics remain practically unaltered under diverse hydroclimatic forcing while the directionality inherent in sediment transport over the hillslope due to gravity reduces the pit counts as the landscape advances towards maturity. To support this reasoning, we analyzed the slope distribution tails for the selected DEMs, which tend to be heavier young landscapes compared to the mature ranges. The relationship between $\Delta{\rm D}_{\rm KR}$ and the fatness of the slope-distribution tails determines the preferential reduction of pits along with smoother hillslopes.

These results have notable implications for the investigations concerning the landscape evolution modeling as well as hydrologic response studies
\cite{tucker2010modelling,chen2014landscape,bonetti2020channelization,walkercomparison}. In particular, current start-of-the-art models that examine water/sediment transport, transient dynamics, and channel forming instabilities for landscape evolution exclude topographic pits from the model (by removing or filling pits during the initial condition of the simulations) \cite{martz1988catch, garbrecht1997assignment, soille2003carving}. The POP value is well above zero even for old smoothened mountainous landscapes, which indicates a restricted scope of these models on simulating and analyzing the mountainous landscape morphology with local extremes during the relaxation phase. Hence, a genuine scope exists for future hydro-geomorphic modeling efforts that include the presence of local extremes during the water flux movement and landscape evolution \cite{ li2011lidar, callaghan2019computing,barnes2020computing}.

\section*{Acknowledgment}
The authors thank Milad Hooshyar for discussions in the early stage of the research. The authors acknowledge support from the US National Science Foundation (NSF) grants EAR-1331846 and EAR-1338694, Innovation Award - Moore Science-to-Action Fund, and BP through the Carbon Mitigation Initiative (CMI) at Princeton University. The authors also acknowledge the support from the High Meadows Environmental Institute (HMEI).

The authors are pleased to acknowledge that the statistical analysis presented in this article was performed on computational resources managed and supported by Princeton Research Computing, a consortium of groups including the Princeton Institute for Computational Science and Engineering (PICSciE) and the Office of Information Technology's High Performance Computing Center and Visualization Laboratory at Princeton University.

\bibliographystyle{elsarticle-num}
\bibliography{reference}
\newpage
\begin{center}
\LARGE{\textbf{Supplementary Information}}
\end{center}

\section*{\label{SS1} Analytical expression for the peak distribution}
\begin{figure}[!hbt]
\centering
\includegraphics[width=\linewidth]{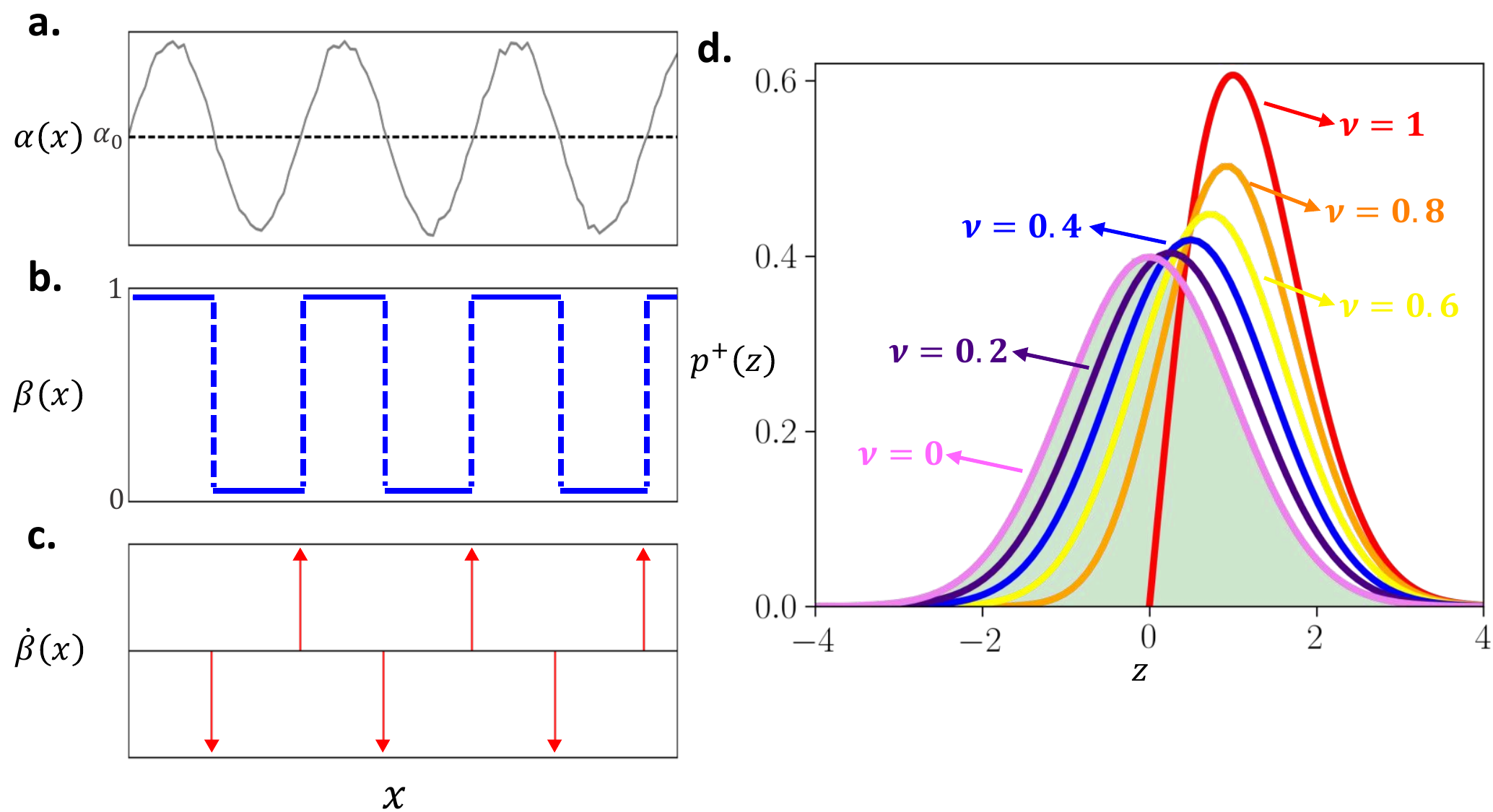}
\caption{(a,b,c): Relationship between excursions and local maxima/peaks. (d): Probability density function for peaks ($p^+(z)$) in a stationary Gaussian stochastic process for different values of $\nu$. Green-colored region shows standard normal distribution.}
\label{fig:Eone}
\end{figure}

The threshold-crossing problem for a mean square differentiable stochastic process $A\left(x\right)$ can be formulated by assuming a new process $B\left(x\right)$ as \cite{soong1973, vanmarcke2010random} 
\begin{equation}
    B(x) = \Theta\left[A(x) -\alpha_o \right],
\label{b_sample_eq}
\end{equation}
which separates the region above and below the threshold $\alpha_o$, where $\Theta[\cdot]$ is the Heaviside step function. Taking the derivative of $B(x)$, we get
\begin{equation}
    \dot{B}(x) = \dot{A}(x) \delta \left[A(x) - \alpha_o \right],
\label{bdot_sample_eq}    
\end{equation}
where $\dot{B}(x)$ presents the derivative of $B(x)$ with respect to $x$ and $\delta [\cdot]$ denotes Dirac delta function. $\dot{B}(x)$ consists of impulses where the positive/negative impulse marks an upward (positive slope)/downward (negative slope) excursion with respect to the threshold $\alpha_0$. Figure \ref{fig:Eone}(a) shows a sample function $\alpha(x)$ of $A(x)$ for which panels (b,c) show corresponding sample functions $\beta(x)$ and $\dot{\beta}(x)$.

Using equation (\ref{bdot_sample_eq}), the expected rate of crossing the threshold $\alpha_0$ for stationary process (independent of $x$) can be simply written as
\begin{equation}
    \gamma_x(\alpha_0) = \int_{-\infty}^{\infty}|\dot{\alpha}| p\left(\alpha = \alpha_0; \dot{\alpha} \right) d\dot{\alpha}.
\label{gamma_cros_eq}  
\end{equation}

The crossing with positive slope (up-crossing) is equal to the crossing with negative slope (down-crossing), which means
\begin{equation}
    \gamma_{x_+}(\alpha_0) = \gamma_{x_-}(\alpha_0) = \frac{\gamma_x(\alpha_0)}{2}
\label{gamma_cros_pm_eq}  
\end{equation}

The direct analogy between threshold problem defined above and the problem of finding extremes becomes apparent when we re-define 
\begin{equation}
    B(x) = \Theta\left[\dot{A}(x)\right],
\end{equation}
where we have instead used the derivative of $A(x)$ for the definition of $B(x)$. Taking derivative of $B(x)$ gives simply
\begin{equation}
    \dot{B}(x) = \ddot{A}(x) \delta \left[\dot{A}(x) \right],
\end{equation}
where the positive impulse indicates the presence of pit or local minima, and the negative impulse represents a local peak. If the information about peaks is useful only above a certain threshold $\alpha_0$, $B(x)$ can be modified as
\begin{equation}
    \dot{B}(x) = \ddot{A}(x) \delta \left[\dot{A}(x) \right] \Theta \left[ A(x) - \alpha_0\right].
\label{bdot_sample_ext_eq} 
\end{equation}

Using equation (\ref{bdot_sample_ext_eq}), the expected rate of finding peaks above $\alpha_0$ for a stationary process can be written as
\begin{equation}
    \gamma_{\dot{x}_+}(\alpha_0) = - \int_{-\infty}^{0}
     d\ddot{\alpha} \int_{\alpha_0}^{\infty}  \ddot{\alpha} p\left(\alpha;\dot{\alpha} = 0; \ddot{\alpha} \right) d\alpha,
\end{equation}
where the integral limit for $\ddot{\alpha}$ to be below zero includes points with negative second derivative for the extraction of peaks and the integral limit for $\alpha$ indicates that values of peaks above the specified threshold are of interest. Putting the threshold $\alpha_0$ as $-\infty$ in the above expression includes all the present peaks.

Using the ratio of expectations, the probability density function of peaks ($p^+(z)$) for stationary standard Gaussian process ($z$) can be analytically obtained \cite{vanmarcke2010random} as
\begin{equation}
p^+(z)= \frac{\sqrt{1-{\nu}^2}}{\sqrt{2 \pi \sigma^2}} \exp \left(\frac{-z^2}{2 {\sigma}^2 \left( 1-{\nu}^2 \right) }\right) + \frac{\nu z}{2 {\sigma}^2} \exp \left( \frac{-z^2}{2 {\sigma}^2}\right) \left[ 1 + {\rm erf} \left( \frac{\nu z}{\sigma \sqrt{2\left(1-{\nu}^2 \right)} }\right)\right],
\label{max_the_dis_sp}
\end{equation}
where $Erf \left( . \right)$ is the error function \cite{zill2011advanced} and $\nu$ is the ratio of expected number of zero up-crossing to the expected number of peaks, mathematically written as
\begin{equation}
    \nu = \frac{\gamma_{x_+}(0)}{\gamma_{\dot{x}_+}(-\infty)} =  \sqrt{\frac{{\lambda_2}^2}{\lambda_0 \lambda_4}}.
\label{nu_eq_sp}
\end{equation}
$\lambda_k$ represents the k${}^{\text{th}}$ spectral moments ($ = \int_{-\infty}^{\infty}|\omega|^k \mathbf{S_\omega} d\omega,$ with $\omega$ as the associated frequency and $\mathbf{S_\omega}$ as the spectral density function). Figure \ref{fig:Eone}(d) presents the distribution of local peaks for varying values of $\nu$ using equation (\ref{max_the_dis_sp}). The red curve shows the limit value $\nu =1$ for which peak distribution is the Rayleigh distribution and the violet curve presents the other limit $\nu=0$ for which the peak distribution matches with the standard Gaussian distribution.

\section*{\label{SS2} Covariance structures}

The Gaussian covariance structure is defined as
\begin{equation}
\gamma \left(r\right) = \sigma^2 \left[ 1 - \exp \left( - \left(\frac{s r}{l}\right)^2\right) \right],
\end{equation}
where $\gamma$ is the semi-variogram, $r$ is the lag distance, $\sigma^2$ is the variance, $s = \frac{\sqrt{\pi}}{2}$ is the standard re-scale factor.

The spherical covariance structure is written as
\begin{equation}
\gamma \left(r\right) = \left\{
\begin{array}{ll}
      1- \frac{3}{2} \cdot s \cdot \frac{r}{l} + \frac{1}{2} \left(s \cdot \frac{r}{l} \right)^3 & r< l/s \\
      0 & r\geq l/s, \\
\end{array} \right.
\end{equation}
where $s = 1$ is the standard re-scale factor \cite{webster2007geostatistics,pyrcz2014geostatistical}. The main distinction between these two classes of covariance models is twice differentiability at zero lag distance for the Gaussian covariance model, while the spherical covariance model does not have well-defined first and second derivative at the zero lag distance \cite{oliver2015basic}. We used GSTools python library for the computations regarding variograms \cite{sebastianmuller2021}.

\section*{\label{SS3} Data collection and processing} 
We employed three different categories of Digital Elevation Models (DEMs) to diminish the bias of the particular type of sensor errors and the applied data-processing algorithm on the final results. For Rocky and Smoky mountain ranges (total 6 DEMs), we utilized 3D Elevation Program (3DEP) DEMs with a grid resolution of arc-second, which are referenced with the North American Datum of 1983 (NAD 83), and the elevations values are provided as 32-bit floating-point (in meters) \cite{us20171}. 3DEP DEMs are derived from interpolating and processing bare-earth lidar cloud points to get high-resolution topographic elevation in uniform rectangular tiles. For the remaining locations across the globe (total 18 DEMs), we employed ASTER (Advanced Spaceborne Thermal Emission and Reflection Radiometer) Global Digital Elevation Model Version 3 \cite{asterv3}. ASTER Version 3, released recently in 2019, provides integer elevation value for an arc-second grid resolution and is shown to be a major advance in the completeness of the anomaly removal compared to the previous versions \cite{abrams2020aster}. We used a DEM of grid resolution 5.19 m obtained by the stereo-pairs of Mars Reconnaissance Orbiter (MRO) Context Camera (CTX) images at NASA Pacific Regional Planetary Data Center (PRPDC) \cite{malin2007context}. The area chosen is in the west of the crater Peta in the Erythraeum Chaos region, which has evidence of being formed by surface erosional processes \cite{barnhart2009long,bouley2010characterization}. 24 DEMs around the globe cover about 1000 km${}^2$ area with an approximate 75 km${}^2$ area covered by DEM `MA' from Mars.

\begin{table}[!htb]
\caption{Information about the DEMs around the globe used in the study}
 \centering
\begin{tabular}{|l|l|l|l|l|l|}
\hline
\textbf{Location} & \textbf{DEM} & \multicolumn{2}{c}{\textbf{Center of the location}} & \textbf{Min. elevation} & \textbf{Max. elevation}\\
    &   & Latitude & Longitude & [m] & [m]\\
\hline
Alps & AP1 & 47.83277778 & 8.158888889 & 421	& 1485\\
     & AP2 & 47.40569444 & 8.853194444	& 409 &	1294\\
     & AP3 & 44.3175 & 6.19 & 577 & 2170\\
\hline
Andes & AD1	& -18.93611111 & -65.37166667 & 2121 & 4177\\
      & AD2	& -32.5775 & -69.16055556 & 1418 & 3435\\
      & AD3	& -0.459166667 & -77.73361111 & 1297 & 3762\\
\hline
Ethiopian Highlands & EH1 & 11.63833333 & 37.61625 & 1616 & 2555\\
                    & EH2 & 12.21513889 & 38.00458333 & 1725 & 3112\\
                    & EH3 & 12.36194444 & 39.39375 & 1630 & 3258\\
\hline
Hamersley & HR1	& -22.32875 & 116.4930556 & 199	& 763\\
          & HR2	& -23.20833333 & 117.8220833 & 316 & 900\\
          & HR3 & -22.70555556 & 116.9155556 & 219 & 922\\
\hline
Rockies & RC1	& 39.31415247 & -107.456299 & 1762.488 & 3607.998\\
 & RC2	& 40.15209818 & -107.5864423 & 1968.8387 & 3290.666\\
 & RC3	& 38.65537594 & -107.1671514 & 2230.178	& 3967.9822\\
\hline
Smokies & SM1 & 35.31257149 & -84.04473995 & 255.76219 & 1691.3047\\
    & SM2	& 35.70307207 & -82.8659371 & 433.8368 & 1601.6133\\
    & SM3	& 35.76222222 & -82.35305555 & 435.68826 & 2027.587\\
\hline
Urals & UR1	& 60.31083333 & 58.93972222 & 261 & 794\\
      & UR2	& 58.615 & 58.46902778 & 221	& 634 \\
      & UR3	& 53.35736111 & 57.66472222 & 305 & 905 \\
\hline
Western Ghat & WG1 & 10.985 & 76.72736111 & 54 & 2069\\
             & WG2 & 10.44125 & 76.72875 & 61 & 1604\\
             & WG3 & 10.30555556 & 77.64180556 & 208 & 2219\\
\hline
\end{tabular}
\label{table1}
\end{table}

\begin{figure}[!hbt]
\centering
\includegraphics[width=\linewidth]{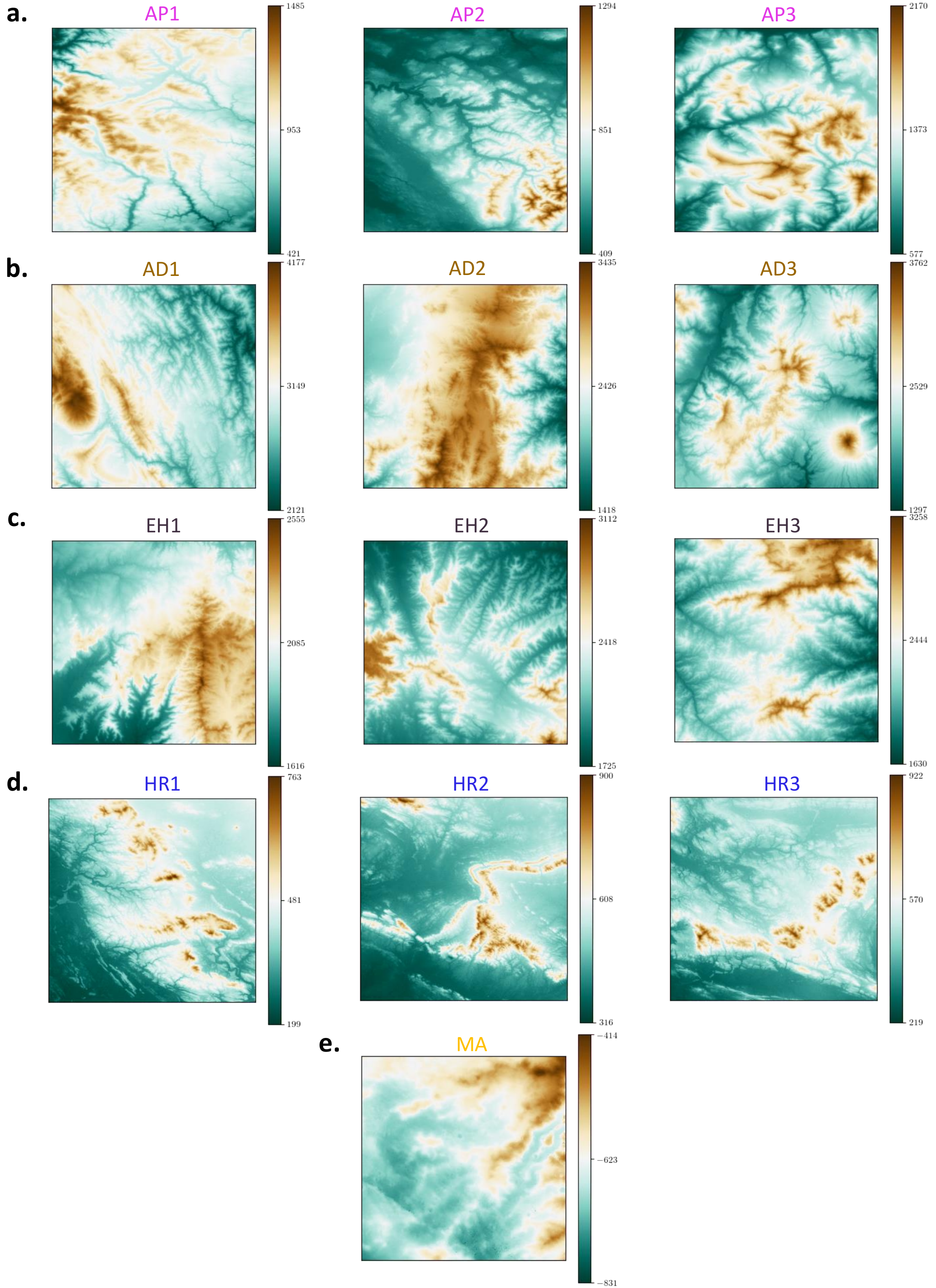}
\caption{DEMs analyzed in this study. (a): Alps, (b): Andes, (c): Ethiopian Highlands, (d): Hamersley Range, and (e): Erythraeum Chaos region (Mars). Elevation values are shown in meter.}
\label{fig:DEM1}
\end{figure}

\begin{figure}[!hbt]
\centering
\includegraphics[width=\linewidth]{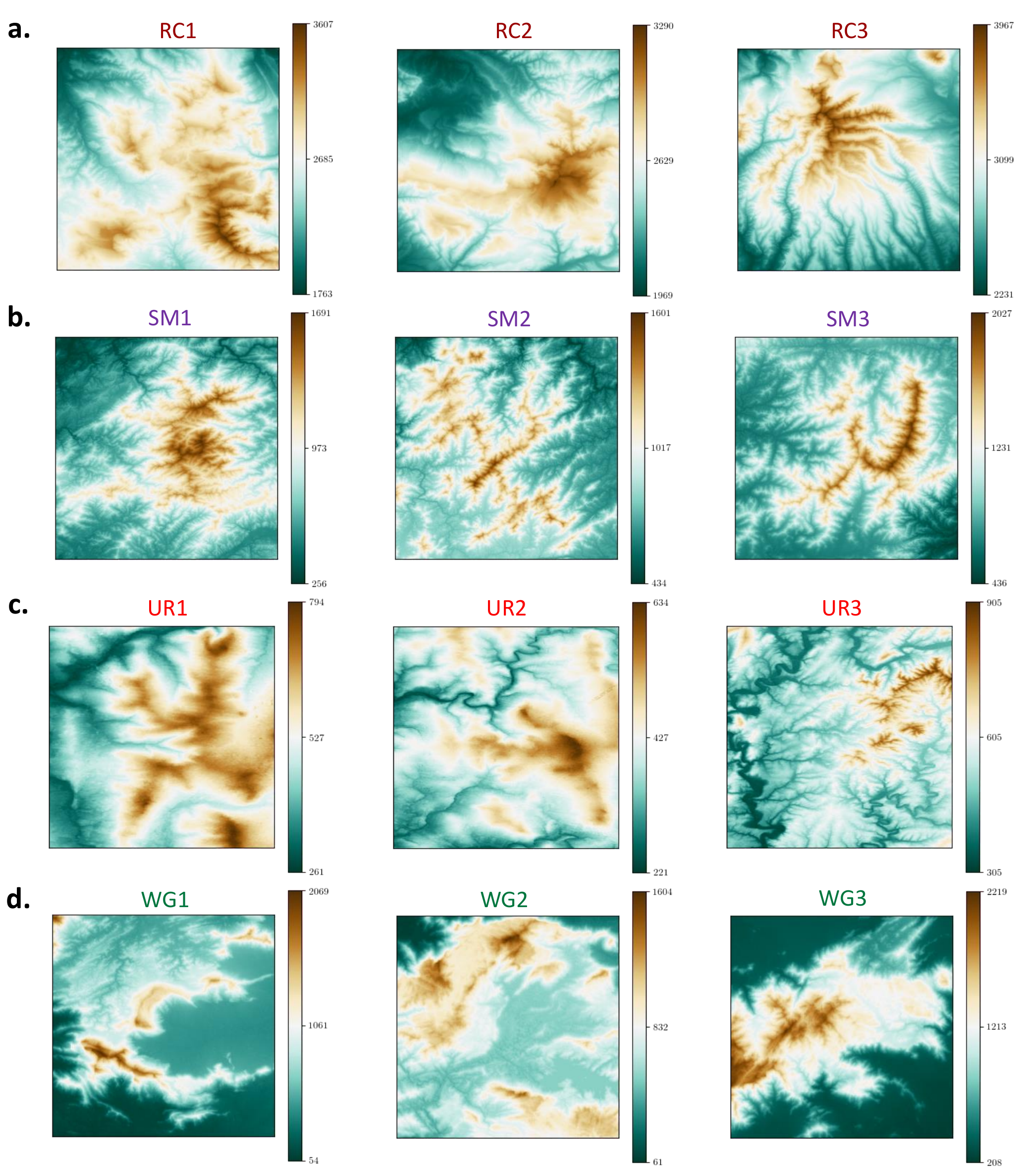}
\caption{DEMs analyzed in this study. (a): Rockies, (b): Smokies, (c): Urals, and (d): Western Ghats. Elevation values are shown in meter.}
\label{fig:DEM2}
\end{figure}

Table \ref{table1} presents the primary information about the DEMs of interest. Before computing the count and frequency distributions of peak and pit, we applied the low-pass Gaussian filter to reduce any possible artificial noise as a pre-processing step, which has been verified to improve the accuracy of DEMs \cite{lashermes2007channel,pelletier2013robust, abdalla2016augmentation}. Figure \ref{fig:DEM1} and \ref{fig:DEM2} present the DEMs used in this study with green color indicating points at low elevation and brown color indicating points at the high elevation in the square domains. Elevation values are shown in meters.

\begin{figure}[!hbt]
\centering
\includegraphics[width=\linewidth]{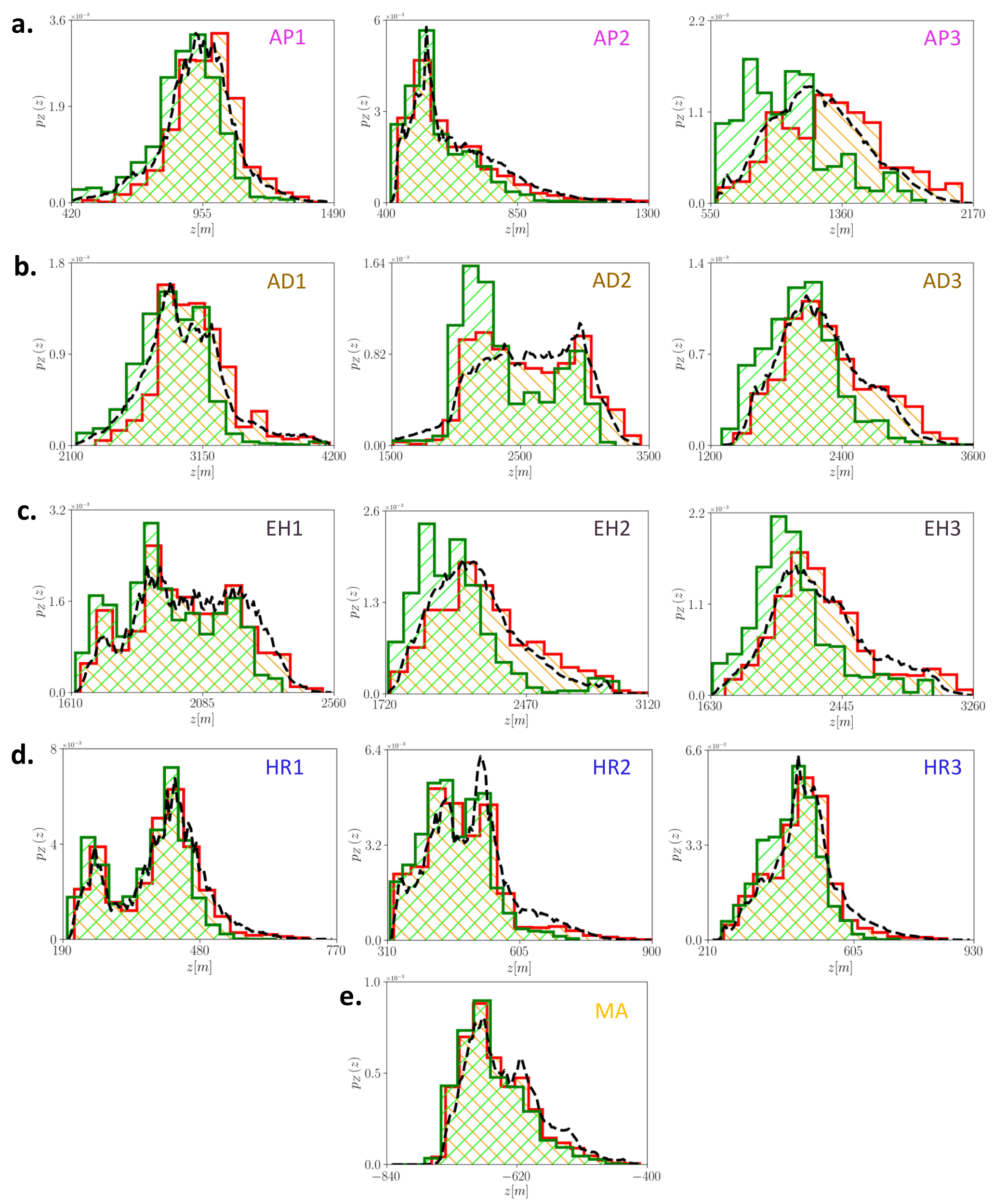}
\caption{Plots of the frequency distributions of elevation field (black-dashed curve), peak (red-hashed step graph), and pit (green-hashed step graph) for (a): Alps, (b): Andes, (c): Ethiopian Highlands, (d): Hamersley Range, and (e): Erythraeum Chaos region (Mars).}
\label{fig:maxmin1}
\end{figure}

\begin{figure}[!hbt]
\centering
\includegraphics[width=\linewidth]{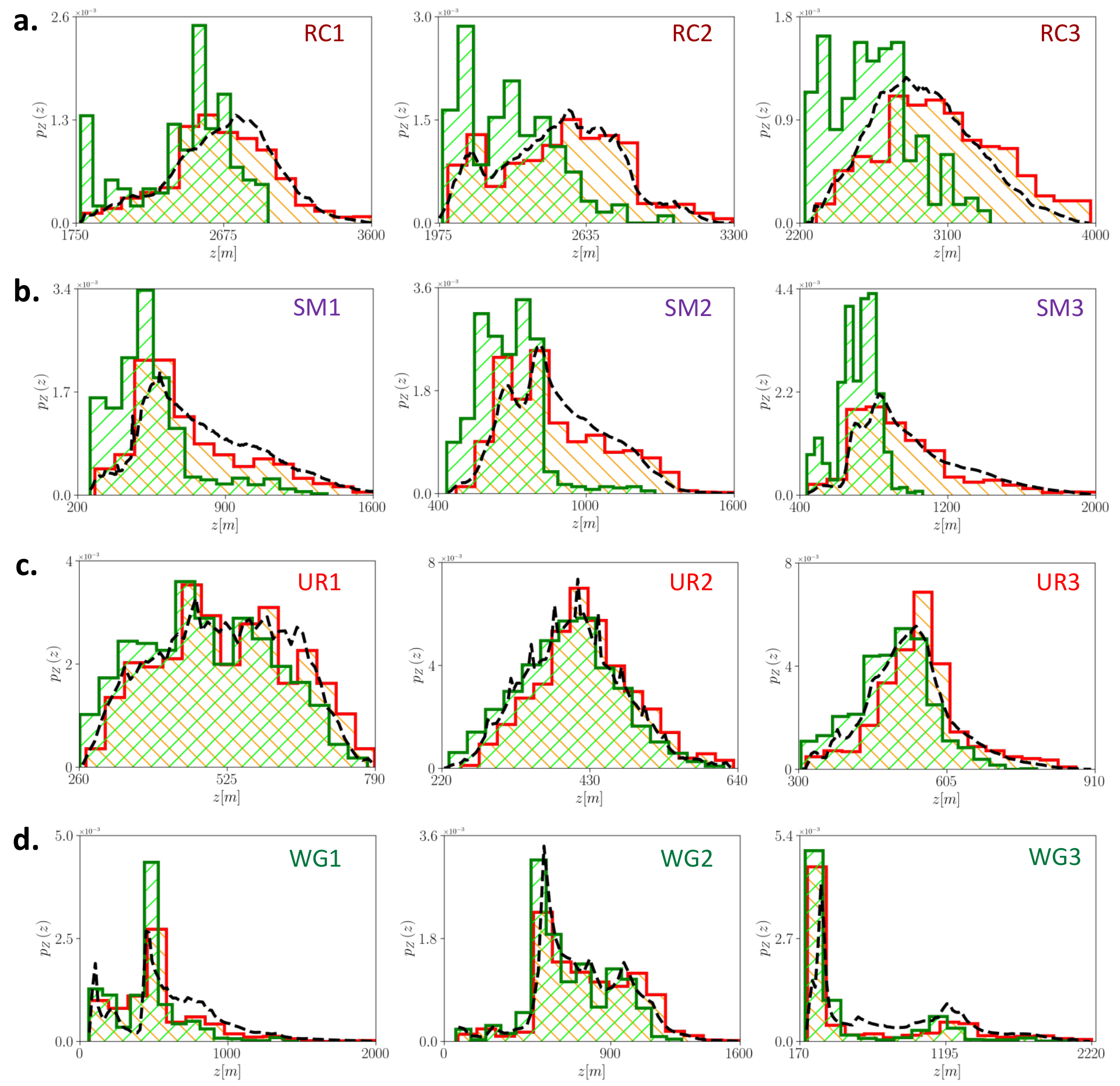}
\caption{Plots of frequency distributions of elevation field (black-dashed curve), peak (red-hashed step graph), and pit (green-hashed step graph) for (a): Rockies, (b): Smokies, (c): Urals, and (d): Western Ghats.}
\label{fig:maxmin2}
\end{figure}

After performing the pre-processing step, the nodes having elevation values higher/lower than the eight adjacent neighbors in the raster grid were taken as peaks/pits.
Figure \ref{fig:maxmin1} and \ref{fig:maxmin2} display the frequency distributions of local peak (red-hashed graph) and pit (green-hashed graph) for 25 DEMs used in the study compared to the elevation field frequency distribution (black curve).

\section*{\label{SS4} Kantorovich–Rubinstein (KR) distance function} 

Kantorovich–Rubinstein (KR) metric is the measure of distance between two probability distributions on a given metric space $\mathcal{M}$. For two distributions $p_1(x)$ and $p_2(x)$, there exists multiple ways utilizing which one can obtain $p_1(x)$ by relocating different pieces of $p_2(x)$ distribution. There is cost associated with every mapping which is the amount of pile/quantity taken from a position in $p_2(x)$ multiplied by the distance traveled. Conceptually, the KR distance represents the most optimal mapping that molds $p_2(x)$ into $p_1(x)$ using the least cost. Hence, for the relatively similar distributions, the amount of work (KR distance) is smaller compared to the distributions that do not agree well. In 1D distribution cases, this metric gets simplified as the area between cumulative distributions ($P_1(x), P_2(x)$) corresponding to the density distributions ($p_1(x), p_2(x)$) as
\begin{equation}
 {\rm KR}\left(p_1,p_2\right) = \int_R|P_1(x) -P_2(x)|dx.
\label{KR_met}
\end{equation}

For estimating the dissimilarity of pit distribution compared to peak and elevation frequency distributions across different mountain ranges, we first normalized elevation values for each DEM to keep the support of all distributions between 0 and 1. 100 bins for the elevation field and 15 bins for peak and pit were employed to obtain the frequency distributions. We subtracted the KR distance of peak and elevation frequency distributions from the KR distance of pit and elevation frequency distributions, which is referred to as $\Delta{\rm D}_{\rm KR}$. This metric provides a quantitative measure of the conformity of pit distributions with elevation and peak frequency distributions.

\section*{\label{SS5} Slope frequency distributions} 

Given a differentiable elevation field $z(x,y)$, the slope is defined analytically as $S= \sqrt{\left(\partial_x z\right)^2+ \left(\partial_y z\right)^2}$. For a raster grid with a certain grid-spacing, both partial derivatives can be obtained numerically using the finite difference method. For all internal nodes, we applied the central-difference method to measure the first-order derivative, which can further be understood as fitting 2nd-degree polynomial to calculate the derivative \cite{langtangen2017finite}. We employed first-order finite-difference for all nodes along the four boundaries. For an internal node z($i,j$) with uniform spacing $\Delta x$ and $\Delta y$ in perpendicular directions, the formulation was
\begin{equation}
    S(i,j) = \sqrt{\left( \frac{z_{i+1,j} - z_{i-1,j}}{2 \Delta x} \right)^2+\left( \frac{z_{i,j+1} - z_{i,j-1}}{2 \Delta y} \right)^2}.
\end{equation}

\begin{figure}[!hbt]
\centering
\includegraphics[width=\linewidth]{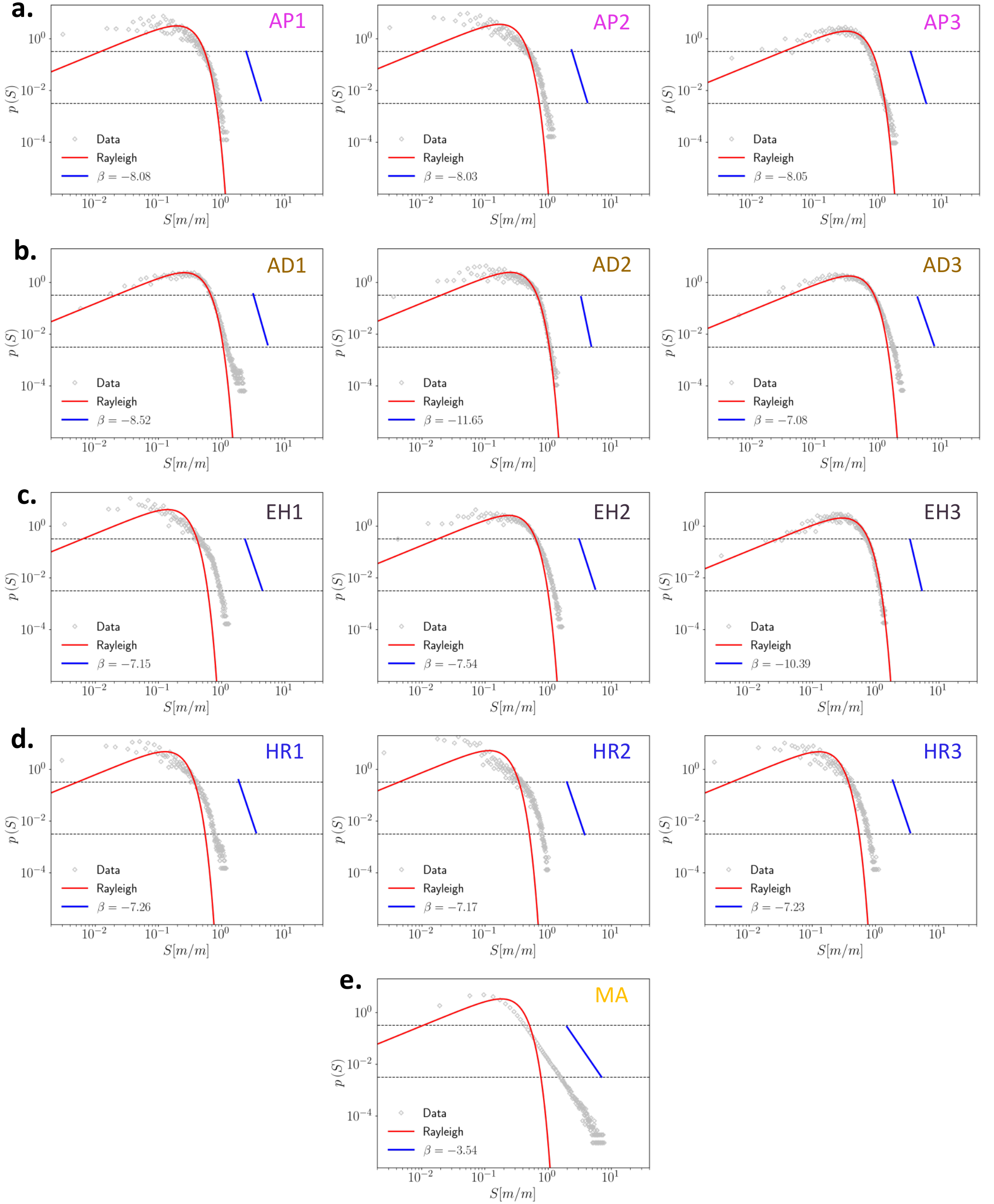}
\caption{Logarithmic plot of the (gray) slope distributions compared to the Rayleigh distributions (red curve). Gray dahsed lines indicate the range of power law fit in the tail from $p(S) = 10^{-0.5}$ to $p(S) = 10^{-2.5}$. Blue line indicates the linear approximation using which the value of $\beta$ is computed. (a): Alps, (b): Andes, (c): Ethiopian Highlands, (d): Hamersley Range, and (e): Erythraeum Chaos region (Mars).}
\label{fig:Slope1}
\end{figure}

\begin{figure}[!hbt]
\centering
\includegraphics[width=\linewidth]{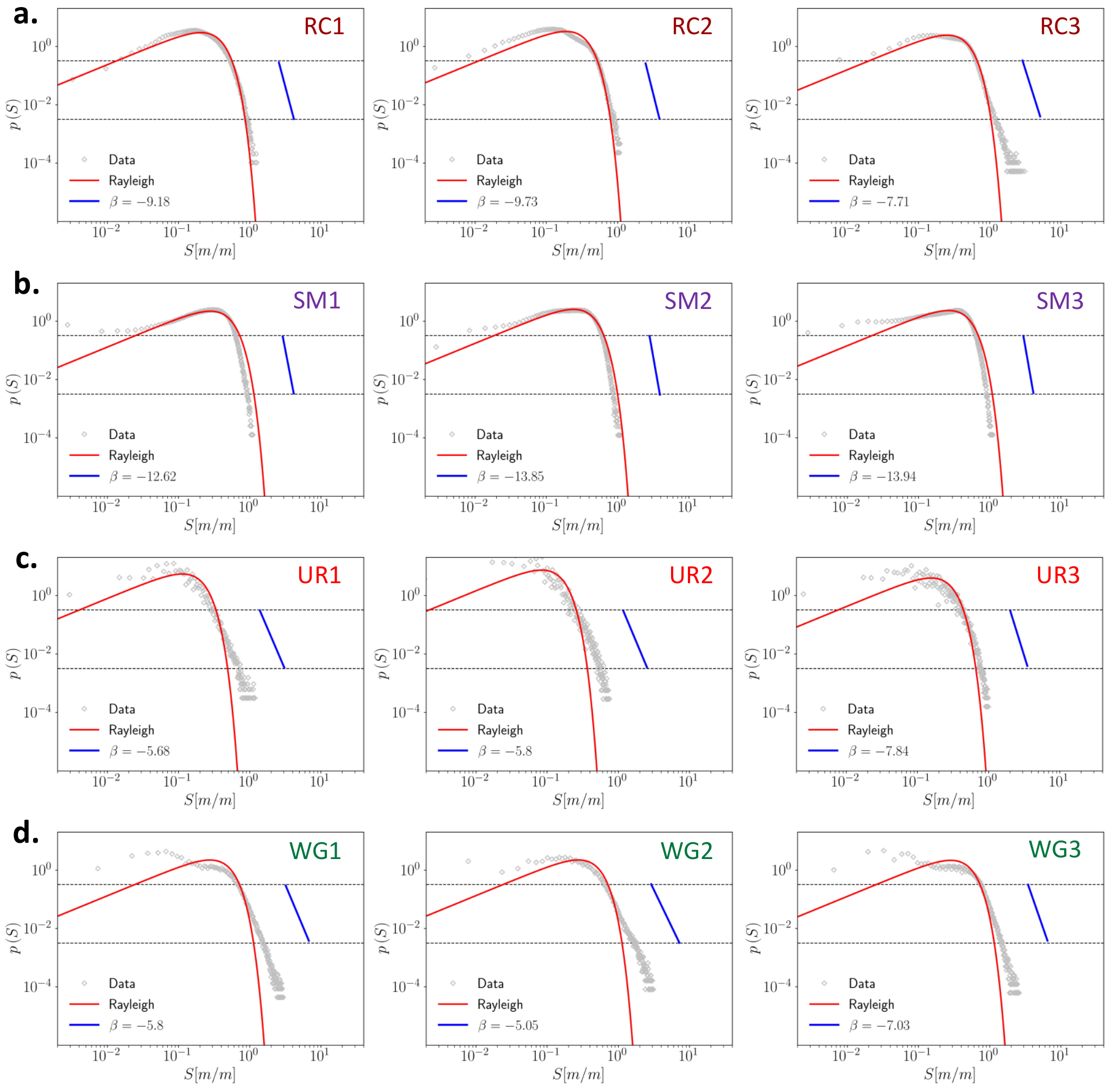}
\caption{Logarithmic plot of the (gray) slope distributions compared to the Rayleigh distributions (red curve). Gray dahsed lines indicate the range of power law fit in the tail from $p(S) = 10^{-0.5}$ to $p(S) = 10^{-2.5}$. Blue line indicates the linear approximation using which the value of $\beta$ is computed. (a): Rockies, (b): Smokies, (c): Urals, and (d):Western Ghats.}
\label{fig:Slope2}
\end{figure}

Figure \ref{fig:Slope1} and \ref{fig:Slope2} show the plots of the slope frequency distributions for selected DEMs in the study (200 bins used for the analysis). The Rayleigh distribution fit is shown by the red curve to visualize the flatness of the observed slope tails \cite{bonetti2017dynamic}. The plot is shown on logarithmic scale with gray dashed lines indicating the range of power law fit in the tail from $p(S) = 10^{-0.5}$ to $p(S) = 10^{-2.5}$ and blue line showing the linear approximation using which the value of $\beta$ is computed. Table \ref{table2} provides the value of POP, $\Delta{\rm D}_{{\rm KR}}$, and $\beta$ along with slope frequency distribution parameters for the DEMs analyzed in this study.

\begin{table}[!htb]
\caption{Information about the mean ($\mu_S$), standard deviation ($\sigma_S$), power-law exponent ($\beta$) of the slope distribution along with the values of POP and $\Delta{\rm D}_{{\rm KR}}$ for the selected DEMs in the study.}
 \centering
\begin{tabular}{|c|c|c|c|c|c|}
\hline
DEM &  $\mu_S$   & $\sigma_S$ &  $\beta$ & POP & $\Delta{\rm D}_{{\rm KR}}$\\
    & (m m${}^{-1}$)    & (m m${}^{-1}$) &   -     & - & -  \\
\hline
\hline
AP1 & 0.190 & 0.137 & -8.080 & 1.097 & 0.0158\\
HR1	& 0.126	& 0.106 & -7.258 & 1.061 & 0.0455\\
HR3 & 0.133	& 0.106 & -7.225 & 1.044 & 0.0401\\
UR3	& 0.156	& 0.108 & -7.840 & 1.013 & 0.0253\\
HR2	& 0.106	& 0.108 & -7.170 & 1.011 & 0.0325\\
MA  & 0.175	& 0.178 & -3.543 & 1.011 & 0.0219\\
WG2 & 0.281	& 0.211 & -5.050 & 0.993 & 0.0222\\
WG1 & 0.262	& 0.230 & -5.801 & 0.974 & 0.0375\\
AD3	& 0.414	& 0.257 & -7.081 & 0.966 & 0.0515\\
UR2	& 0.082	& 0.058 & -5.800 & 0.962 & -0.0160\\
WG3 & 0.293	& 0.243 & -7.030 & 0.919 & 0.0616\\
AP2 & 0.156 & 0.137 & -8.033 & 0.913 & 0.0523\\
AD1 & 0.330 & 0.185 & -8.516 & 0.911 & 0.0225\\
UR1	& 0.116	& 0.074 & -5.680 & 0.897 & 0.0436\\
EH1 & 0.150	& 0.131 & -7.150 & 0.831 & 0.0672\\
EH2 & 0.268	& 0.189 & -7.537 & 0.780 & 0.0475\\
AP3 & 0.342	& 0.189 & -8.049 & 0.747 & 0.0756\\
EH3 & 0.343	& 0.185 &-10.390 & 0.698 & 0.0901\\
SM1 & 0.319	& 0.152 & -12.620 & 0.554 & 0.1325\\
AD2 & 0.281 & 0.189 & -11.650 & 0.474 & 0.0690\\
SM2	& 0.285	& 0.149 & -13.850 & 0.408 & 0.1460\\
SM3	& 0.316	& 0.161 & -13.940 & 0.388 & 0.1494\\
RC2	& 0.206	& 0.132 & -9.734 & 0.302 & 0.1501\\
RC3	& 0.298	& 0.175 & -7.710 & 0.240& 0.1190\\
RC1	& 0.230	& 0.132 & -9.180 & 0.143 & 0.1194\\
\hline
\end{tabular}
\label{table2}
\end{table}

\end{document}